# Structure and composition of C-S-H compounds up to 143 GPa

Elena Bykova[1], Maxim Bykov[1,2], Stella Chariton[3], Vitali B. Prakapenka[2], Konstantin Glazyrin[4], Andrey Aslandukov[5], Alena Aslandukova[6], Giacomo Criniti[6], Alexander Kurnosov[6], Alexander F. Goncharov[1]

[1] Earth and Planets Laboratory, Carnegie Institution of Washington, 5251 Broad Branch Road NW, Washington, DC 20015, USA

[2] Department of Mathematics, Howard University, Washington, DC 20059, USA

[3] Center for Advanced Radiations Sources, University of Chicago, Chicago, Illinois 60637, USA

[4] Photon Science, Deutsches Elektronen-Synchrotron, Notkestrasse 85, 22607 Hamburg, Germany

[5] Material Physics and Technology at Extreme Conditions, Laboratory of Crystallography, University of Bayreuth, D-95440 Bayreuth, Germany

[6] Bavarian Research Institute of Experimental Geochemistry and Geophysics (BGI), University of Bayreuth, Univesitaetsstrasse 30, 95440 Bayreuth, Germany

**We synthesized two C-S-H compounds from a mixture of carbon and sulfur in hydrogen and from sulfur in mixed methane-hydrogen fluids at 4 GPa. X-ray synchrotron single-crystal diffraction and Raman spectroscopy have been applied to these samples up to 58 and 143 GPa, respectively. Both samples show a similar Al$_2$Cu type *I*4/*mcm* basic symmetry, while the hydrogen subsystem evolves with pressure via variously ordered molecular and extended modifications. The methane bearing sample lowers symmetry to an orthorhombic *Pnma* structure after laser heating to 1400 K at 143 GPa. The results suggest that superconducting C-S-H compounds are structurally different from a common *Im*-3*m* H$_3$S.**

The concept of doped hydrogen metallic alloys [1] has been reinvigorated in the last several years fueled by developments of high-pressure experimental techniques and more advanced computational methods. Recent discoveries of high-temperature superconductivity in poly- and super-hydrides [2-7] at high pressures (100-270 GPa) call for investigations of the properties related to its emergence. However, the experiments at such extreme pressures remain very challenging especially in combining different experimental techniques often making uncertain the conclusions about the nature of superconductivity (*e.g.* Ref. [8]). In particular, the direct structural determination including the hydrogen positions remains problematic and is commonly addressed using first principles density functional theoretical calculations (*e.g.* Ref. [9]). Also, the superconductivity probes are often not fully convincing as the direct measurements of the Meissner effect are extremely challenging [10]. Theoretical models mostly lean toward strong electron-phonon coupling mechanisms, which involve high-frequency hydrogen phonon modes[11, 12], where anharmonicity plays an important role as well the hydrogen crystallographic positions and dynamics [13]. However, these theories are challenged by alternative or more complex theoretical descriptions (*e.g.* Refs. [14, 15]). Moreover, the whole notion of high-temperature superconductivity is questioned because of the alleged inconsistencies in the superconducting probes [16].

Historically, sulfur polyhydride, $H_3S$ was the first material, which demonstrated a very high superconducting transition temperature $T_c$ of 203 K at 150 GPa [2]. This material has an unusual for ambient pressure composition and can be considered as being disproportionated from a familiar molecular $H_2S$ [17]. However, at low pressures a host-guest compound $(H_2S)_2H_2$ with the $Al_2Cu$ structure (*I4/mcm* symmetry) has been found, which has the same stoichiometry as $H_3S$ high-temperature superconductor [18]. The pressure driven transformation pathway between these two phases has been demonstrated on the pressure decrease upon temperature annealing to overcome the kinetic hindrance [19], while on the compression path *I4/mcm* or *Cccm* (at higher pressures) $(H_2S)_2H_2$ remains stable up to 160 GPa [20, 21]. This is again due to kinetic reasons because *Cccm* $(H_2S)_2H_2$ and *Im-3m* $H_3S$ are very different structurally, and the transition between them involves the volume discontinuity of more than 4% (at 110 GPa) [19]. However, *Im-3m* $H_3S$ has been reported to form above 140 GPa from the elements or reaction products of $H_2S$ by temperature cycling below 300 K [22] or with the assistance of gentle laser heating [19, 23] (<1300 K).

Recently, Sneider *et al.*[7] reported much higher values of $T_c$ in a mixed C/S polyhydride compressed up to 267 GPa. Based on Raman spectroscopy observations, the synthesized at 4 GPa material was suggested to be the $(CH_4)_x(H_2S)_{(2-x)}H_2$ compound with *I4/mcm* symmetry [18], but no direct structural determination was presented. The substitution of $H_2S$ by methane is plausible because both molecules have similar kinetic diameters; however, the carbon composition of the material they synthesized was not determined. Moreover, since no Raman signal has been recorded at pressures above 60 GPa, the structure of the superconducting phase could not be determined or traced. Theoretical calculations on this system [24, 25] showed dynamic stability of a mixed $CSH_7$ hydride above 100 GPa, where $CH_4$ molecules substitute $SH_6$ units in *Im-3m* $H_3S$ causing a lattice distortion. This material is found to be a high-temperature superconductor comparable to $H_3S$ but no sharp increase of $T_c$ with pressure as reported by Sneider *et al.*[7] is predicted. Here we investigated two C-S-H materials referred hereafter as B1 and B2 (see details in Supplemental Material [26]) synthesized similarly to Ref.[7] with elementary carbon and methane used as the carbon sources and explored their structure, vibrational and optical properties using single-crystal X-ray diffraction (SCXRD), Raman and visible absorption spectroscopy, concomitantly up to 143 GPa. We determined that the structures of these alloys are very similar, and they resemble that of *I4/mcm* $(H_2S)_2H_2$, which slightly distorts but retains the main structural motif even after laser heating to beyond 2400 K at 143 GPa. This suggests that carbon stabilizes $Al_2Cu$ type structure, which is responsible for a great increase in superconductivity at high pressures.

We examined four samples of two different kinds at various pressures by SCXRD at GSECARS (APS, Chicago) and ECB at Petra-III (DESY, Hamburg). The details of the sample preparation, data collection, structure determination and refinement are presented in the Supplemental Material [26]. The exact determination of hydrogen positions attached to S atoms was not possible, due to low scattering power of hydrogen compared to sulfur and likely due to strong degree of disorder in $H_2S$ molecules (see below). On the other hand, the location of hydrogen molecules could be determined by analyzing residual density maps. Hydrogen molecules were introduced in the structure in order to describe the highest residual density reflections, additionally we used constraints on the H-H distances (0.75±0.02 Å according to Duan *et al.*[27]); this procedure improved $R_1$ values by 0.5-1 percent points. For the B1 and B2 samples, the substitution of sulfur by carbon does not improve the refinement, since the basic *I4/mcm* structure has only one crystallographic position for S atoms, which can be randomly substituted by C atoms, and its refined occupancy is directly correlated with a scale factor.

The basic structure of the material is defined solely by sulfur positions and belongs to tetragonal *I4/mcm* $(H_2S)_2H_2$[18] type. We find that below 14 GPa $H_2$ molecules are oriented parallel to *c*-axis and located in the cavities formed by sulfur (Fig. 1(b)). The symmetry positions of S atoms (Tables S2-S4 [26]) remain unchanged while the fractional coordinates change little to the highest studied pressure (143 GPa). The reciprocal space reconstructions (Fig. 1(a)) demonstrate that neither systematic absence violations for *I*-lattice nor new reflections appear, and the positions of the diffraction peaks show no splitting. Also, we find no orthorhombic distortion (*e.g.* in *Cccm* structure as suggested in Ref. [27]) of the basic structure up to the highest reached pressures for both B1 and B2 materials. However, in the B1 batch sample at 27 GPa additional reflections were observed nearby the reflections belonging to the *I4/mcm* phase. The new phase was indexed in monoclinic *C*-centered unit cell with parameters $a = 8.255(16)$, $b = 6.3944(18)$, $c = 12.14(3)$ Å, $\beta = 99.3(3)°$ (Table S5 [26]). This corresponds to the tripling of the *I4/mcm* unit cell tentatively accounting for very complex Raman spectra observed in this pressure range (see below).

Under compression above 14 GPa the refinement agreement factors are systematically better if the $H_2$ molecules initially oriented along the *c*-axis are rotated by 90º so they become parallel to the *ab* plane. In the refinement such molecules were approximated by two perpendicular molecules with occupancy factor of 0.5 to keep the four-fold axis (Fig. 1(b)). No further changes in the orientation of $H_2$ molecules were detected up to 143 GPa.

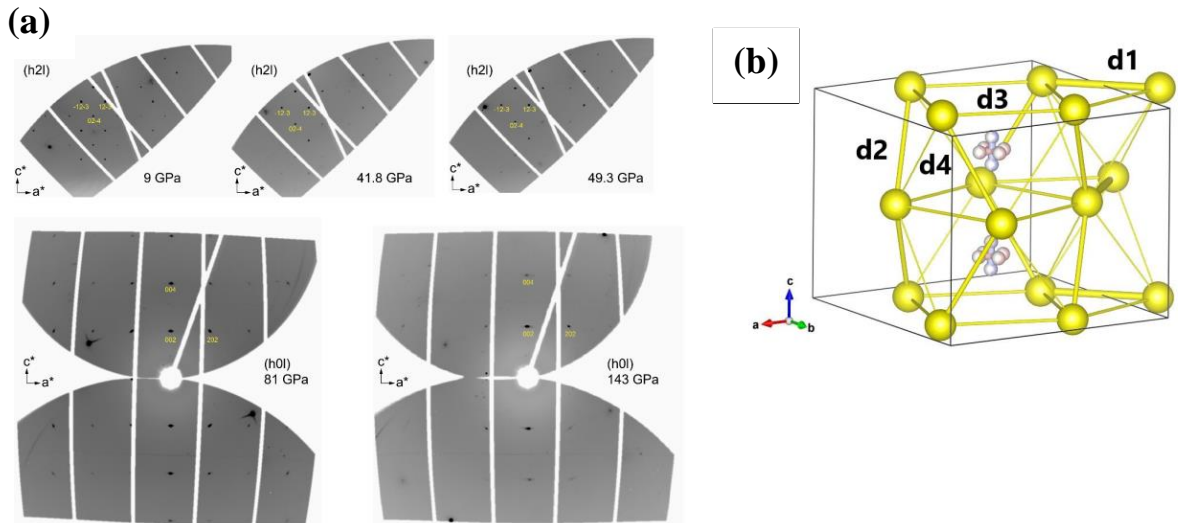

**Figure 1**. XRD results on C-S-H compounds at high pressures. (a) Reconstructed reciprocal lattice plane defined by [100] and [001] directions of methane substituted B2 batch samples at various pressures. (b) the structural model of *I4/mcm* $(H_2S)_2H_2$ (see SM for the unit cell parameters and atomic positions). Only the $H_2$ molecules were included to the structural model: the oriented along the *c* axis and in the *ab* plane below and above 14 GPa- both are shown. The S-S closest distances

are labeled. Following laser heating at 143 GPa, we find a distortion of the unit cell, which becomes orthorhombic *Pnma*, but the arrangement of sulfur atoms repeats one in the parent *I4/mcm* structure; the positions of hydrogen atoms could not be determined. The potential positions of hydrogen atoms forming hydrogen bonds in the S cages can be inferred by analyzing the P dependencies of the d1 to d4 (Fig. 2).

While the location of H atoms in S(C) cages was not possible, the shortest S(C)-S(C) distances deduced from our single-crystal XRD data (Fig. 2(a)) could be used to infer their positions. The S(C)-S(C) distances in the samples of both types contract similarly under pressure swapping the ranking between d1 and d2 and splitting d3 and d4. At about 50 GPa, where the Raman spectra dramatically change signaling the formation of an extended solid (see below), the d3 and d4 distances are in the range of 3.13-3.2 Å, corresponding to a regime of the strong hydrogen bonding near symmetrization [27]. The d1 and d2 distances are shorter that one would accept if there is an H atom between them forming a nearly linear hydrogen bond. These results suggest that H atoms in the S(C) cages are likely positioned along the lines corresponding d3 and d4 S(C)-S(C) distances.

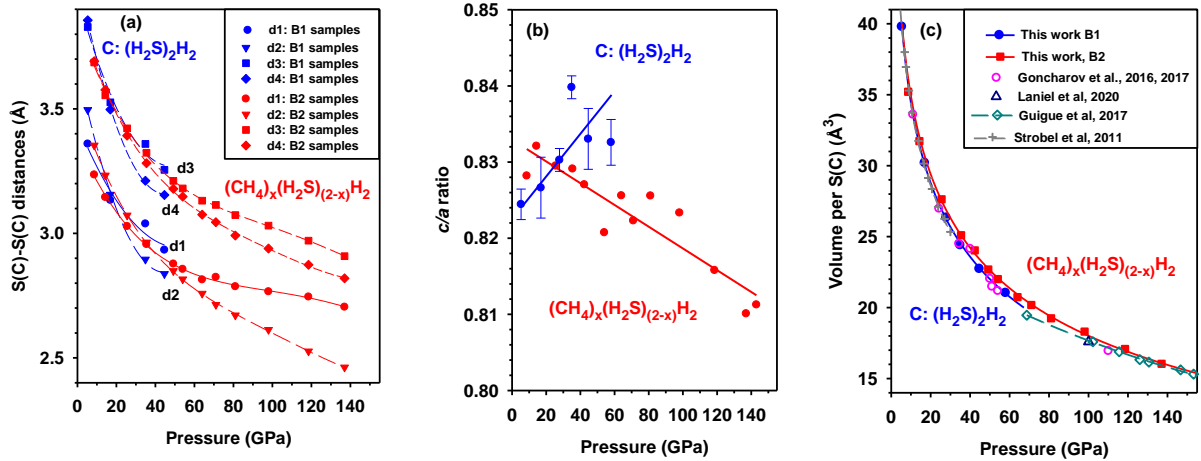

**Figure 2**. The unit cell and structure parameters determined in single-crystal XRD experiments The result for both B1 and B2 batches are shown in blue and red, respectively. (a) The closest S(C)-S(C) distances; (b) The *c/a* lattice parameters ratio; (c) The volumes per one S(C) atom; the results of this work for batches B1 and B2 (carbon and methane originated) are fit with the Vinet EOS. The error bars are smaller than the symbol size. Our data are compared to those of Refs. [17-20, 28] for *Cccm* (*I4/mcm*) $(H_2S)_2H_2$; the results of Pace *et al*. [21], which agree well with other data at high pressures but scatter and largely disagree below 60 GPa, are not shown for clarity.

There is a difference in the behavior of samples from batches B1 and B2 which can be assigned to a different kind of substitution of S by C. In the samples of the batch B1, which do not contain the methane as a building block, carbon (if any) likely substitutes $H_2S$ molecules- we call it C doped $(H_2S)_2H_2$. In the samples of batch B2, which do contain methane, it supposedly substitutes $H_2S$ molecule forming $(CH_4)_x(H_2S)_{(2-x)}H_2$. It is interesting that B1 and B2 samples demonstrate a distinct compression anisotropy (Fig. 2(b)): the methane containing samples compress stronger along the *c* axis, while carbon doped samples are more compressible along *a* and *b*. At nearly 4

GPa, where they were synthesized, the samples have indistinguishable lattice parameters. Under pressure, a systematic difference in volumes for batches B1 and B2 develops. The samples synthesized from S:CH$_4$:H$_2$ mixtures have larger volumes, while those synthesized from C:S:H$_2$ have the EOS, which is very close to that of (H$_2$S)$_2$H$_2$ synthesized by various techniques at diverse P-T conditions (Fig. 2(c)). The synthesized here B1 material seems to have slightly larger volume in the limit of high pressures, but the difference is within the experimental uncertainties. It is interesting that the EOS of B2 samples merges with that of pure (H$_2$S)$_2$H$_2$ at high pressures. While the experimental EOS of (H$_2$S)$_2$H$_2$ has been shown to be in a good agreement with theoretical calculations, no theoretical results has been reported for mixed (CH$_4$)$_x$(H$_2$S)$_{(2-x)}$H$_2$ compounds with *I*4/*mcm* or *Cccm* structure.

At the highest reached pressure of 143 GPa, the B2 sample was laser heated twice (both at GSECARS and ECB at Petra-III) up to a maximum 2700 K. Already after the first heating the sample re-crystallized in a multi-grain phase. The most intense grain was indexed in an orthorhombic *P*-centered unit cell with parameters $a = 7.591(5)$, $b = 4.439(13)$, $c = 7.641(6)$ Å Table S6 [26], Fig. S2). The unit cell volume is doubled compared to the parent *I*4/*mcm* structure. Indeed, the reciprocal space reconstructions show appearance of the reflections violating *I*-lattice symmetry and additional superlattice reflections at 0.5($a$*+ $b$*). The best structure solution has been achieved within SG *Pnma*, where only positions of S atoms can be located. The interatomic S…S distances vary widely from to 2.401(7) to 3.004(6) Å, but general arrangement of the heavy atoms still repeats one of the *I*4/*mcm* structure. Such distances would also suggest strong hydrogen bonding in the structure. The increase in the unit cell volume by 1.7(2) Å$^3$ compared to *I*4/*mcm* volume before heating suggests that this distorted structure could have incorporated one extra H atom (*e.g.* Ref. [29]) yielding (CH$_4$)$_x$(H$_{2+\delta}$S)$_{(2-x)}$H$_2$, where δ can vary from 0.125 to 0.25 if 0<x<1.

To learn more about the H atom positions and C composition of our samples, we concomitantly investigated Raman spectra of the samples probed by XRD and other samples, which were synthesized similarly. The Raman spectra of B1 samples synthesized in this work (Fig. S3) are very similar to those reported in Ref. [7], except no sharp C-H modes have been detected. They are also similar to Raman spectra of (H$_2$S)$_2$H$_2$ reported previously [18, 19, 21]. This includes the behavior of the lattice, S-H bending and stretching, and H-H stretching modes. Similarly to reported in Ref. [7] for the proposed methane bearing material, these data show a sequence of phase transitions at approximately 15, 35, and 45 GPa manifesting two-stage ordering of H$_2$S molecules and their polymerization via formation of symmetric hydrogen bonds [18, 19, 21]. Our B1 samples show a similar transformation sequence (Fig. S3) suggesting that they contain little if any carbon with an unknown substitution pattern. This is consistent with our structural data and volume vs pressure dependencies (Figs. 1, 2). The Raman spectra above 25 GPa show very complex structure for all types of vibrations. This is consistent with the XRD observations of the superstructure, suggesting that the observed mode splitting is due the Brillouin zone folding, which makes the zone boundary vibrations Raman active (*e.g.* Ref. [30]). Above 45 GPa, the majority of the Raman bands disappear, and only weak broad low-frequency bands remain that can be assigned to S-S modes similar to observations in H$_2$S [17].

In contrast, our B2 samples do show the incorporation of methane molecules into the same lattice manifested by the appearance of red shifted C-H stretching $\nu_1$ and $\nu_3$ modes (Fig. S4, Fig. 3), which can be distinguished from methane molecules incorporated in an inclusion compound with molecular $H_2$ [31]. The major features of the pressure behavior of B2 materials are similar to those for B1 materials but there are important differences. The lattice mode spectrum is much simpler showing just a few broad bands; the S-H and H-H stretching modes have less components, their softening behavior is much less pronounced including the frequency shift and pressure induced peak broadening. Instead, the associated peaks are broadened at the transition to a state, where $H_2S$ molecules are expected to cease rotations (above 15 GPa), and likely form static orientationally disordered state. As in the case of B1 samples, the Raman spectra of methane bearing B2 samples change drastically above 50 GPa with all the major peak vanishing and only a few lattice modes preserving to higher pressures. This indicates the transformation of this material into an extended state.

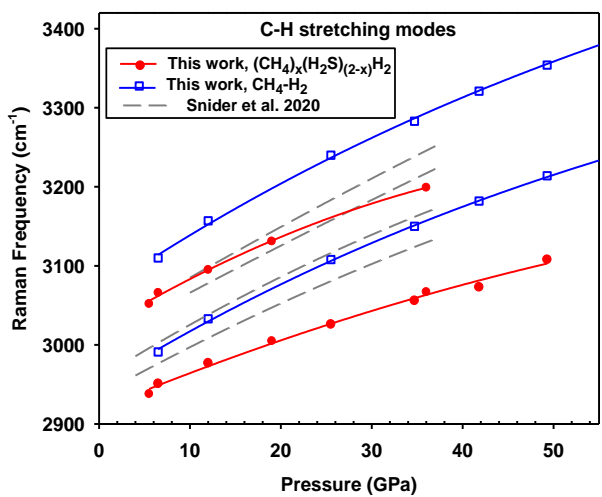

**Figure 3**. Pressure dependence of the C-H Raman modes measured here on the B2 samples in comparison with the results of Snider *et al.* [7].

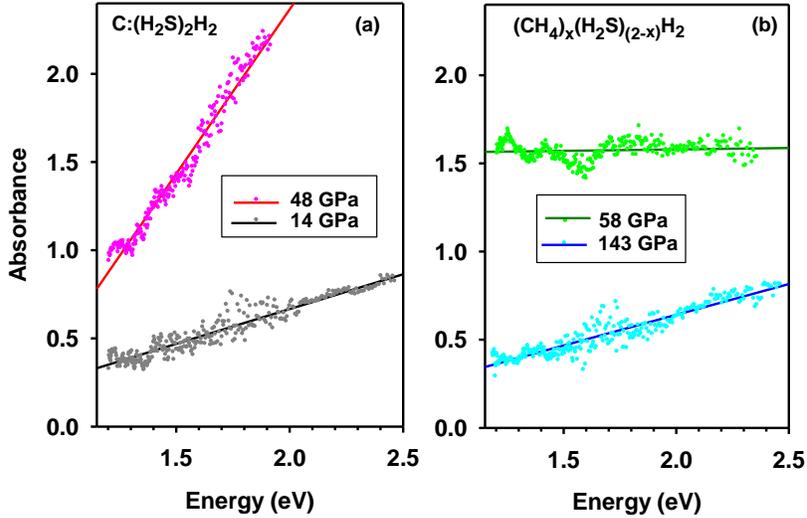

**Figure 4**. Optical absorption spectra at various pressure of (a) B1 C doped $(H_2S)_2H_2$ and (b) B2 $(CH_4)_x(H_2S)_{(2-x)}H_2$. The B2 sample at 143 GPa is probed after the laser heating experiment. The spectra are normalized to the transmission of the medium — $H_2$ and mixed $H_2$-$CH_4$ medium, respectively.

The optical properties of both materials synthesized here change under pressure. While they are translucent at low pressure (Fig. S1), they gradually become dark and transmit little in extended high pressures states. The absorption spectra of B1 C doped $(H_2S)_2H_2$ samples show a broad absorption edge, which shifts to lower energies at high pressures, but the sample remained a narrow band semiconductor up to the highest pressure reached of 58 GPa (Fig. 4). In contrast, B2 samples in an extended state at 58 GPa shows a featureless absorption suggesting indirect gap closure or semimetal behavior. It is interesting that the sample of the same batch measured after the laser heating at 143 GPa opens a bandgap, likely due to hydrogen ordering.

Our combined XRD, Raman, and optical spectroscopy up to 143 GPa provides missing previously information about the composition, and pressure dependent structure and electronic properties of the reported room temperature superconducting C-S-H compounds. The composition and thus properties strongly depend on the synthesis conditions and procedure. Our attempts to reproduce synthesis of $(CH_4)_x(H_2S)_{(2-x)}H_2$ compounds from a mixture of C and S in fluid $H_2$ medium [7] were unsuccessful in that the materials we obtained did not show any sign of C-H bonds. We argue that the materials reported by Snider *et al.* [7] did not incorporate methane molecules as they infer. Indeed, our B2 samples synthesized in a mixed $H_2$ and $CH_4$ gas do show such incorporation beyond any doubt, as both XRD and Raman provide clear evidence for such mixed crystal (Figs. 1, 2, S2). However, these samples show clearly different Raman spectra compared to those reported in Ref. [7] (Fig. S3). Moreover the comparison of the pressure dependencies of the Raman frequencies of the C-H stretching modes (Fig. 3) shows the difference between our data and those of Snider *et al.* [7], suggesting that their C-H bands do not belong from to material of interest. We speculate that they originate from the CH$_4$-H$_2$ compounds [31] that have been synthesized in the cavity during the "photochemistry" processes.

In conclusion, using synchrotron single-crystal XRD technique we determined the basic symmetry of C-S-H compounds synthesized in the DAC in the conditions that are similar to those reported recently for room temperature superconductor above 140 GPa. The sulfur lattice forms an Al$_2$Cu structure (*I*4/*mcm* symmetry), which remains to the highest pressure of 143 GPa reached here. The superstructure and lattice distortion detected at 27 GPa signals the formation of a hydrogen ordered state. Raman spectra show a series of molecular ordering transformations in the hydrogen subsystem, which eventually results in formation of extended narrow-gap semiconducting or semimetallic solids above 60 GPa. The stability of the basic sulfur-based cage with *I*4/*mcm* symmetry up to 143 GPa suggests that this structure may be a host of room temperature superconducting state unlike *Im*-3*m* H$_3$S revealing superconductivity at 203 K at 150 GPa.

Parts of this research were carried out at the GeoSoilEnviroCARS (The University of Chicago, Sector 13), Advanced Photon Source, Argonne National Laboratory. GeoSoilEnviroCARS is supported by the National Science Foundation - Earth Sciences (EAR - 1634415) and Department of Energy GeoSciences (DE-FG02-94ER14466). The Advanced Photon Source is a U.S. Department of Energy (DOE) Office of Science User Facility operated for the DOE Office of Science by Argonne National Laboratory under Contract No. DE-AC02-06CH11357. Parts of this research were carried out at the Extreme Conditions Beamline (P02.2) at DESY, a member of Helmholtz Association. We acknowledge support by the Army Research Office and was accomplished under the Cooperative Agreement Number W911NF-19-2-0172 and Carnegie Institution of Washington.

1.	N. W. Ashcroft, Physical Review Letters **92** (18), 187002 (2004).
2.	A. P. Drozdov, M. I. Eremets, I. A. Troyan, V. Ksenofontov and S. I. Shylin, Nature **525** (7567), 73-76 (2015).
3.	A. P. Drozdov, P. P. Kong, V. S. Minkov, S. P. Besedin, M. A. Kuzovnikov, S. Mozaffari, L. Balicas, F. F. Balakirev, D. E. Graf, V. B. Prakapenka, E. Greenberg, D. A. Knyazev, M. Tkacz and M. I. Eremets, Nature **569** (7757), 528-531 (2019).
4.	M. Somayazulu, M. Ahart, A. K. Mishra, Z. M. Geballe, M. Baldini, Y. Meng, V. V. Struzhkin and R. J. Hemley, Physical Review Letters **122** (2), 027001 (2019).
5.	P. P. Kong, V. S. Minkov, M.A.Kuzovnikov, S.P.Besedin, A. P. Drozdov, S. Mozaffari, L. Balicas, F.F. Balakirev, V. B. Prakapenka, E. Greenberg, D. A. Knyazev and M. I. Eremets, ArXiv190910482 Cond-Mat (2019).
6.	D. V. Semenok, A. G. Kvashnin, A. G. Ivanova, V. Svitlyk, V. Y. Fominski, A. V. Sadakov, O. A. Sobolevskiy, V. M. Pudalov, I. A. Troyan and A. R. Oganov, Materials Today **33**, 36-44 (2020).
7.	E. Snider, N. Dasenbrock-Gammon, R. McBride, M. Debessai, H. Vindana, K. Vencatasamy, K. V. Lawler, A. Salamat and R. P. Dias, Nature **586** (7829), 373-377 (2020).
8.	A. Majumdar, J. S. Tse and Y. Yao, Angewandte Chemie International Edition **56** (38), 11390-11393 (2017).

9.  C. M. Pépin, G. Geneste, A. Dewaele, M. Mezouar and P. Loubeyre, Science **357** (6349), 382-385 (2017).
10. X. Huang, X. Wang, D. Duan, B. Sundqvist, X. Li, Y. Huang, H. Yu, F. Li, Q. Zhou, B. Liu and T. Cui, National Science Review **6** (4), 713-718 (2019).
11. K. Tanaka, J. S. Tse and H. Liu, Physical Review B **96** (10), 100502 (2017).
12. I. Errea, M. Calandra, C. J. Pickard, J. Nelson, R. J. Needs, Y. Li, H. Liu, Y. Zhang, Y. Ma and F. Mauri, Physical Review Letters **114** (15), 157004 (2015).
13. I. Errea, M. Calandra, C. J. Pickard, J. R. Nelson, R. J. Needs, Y. Li, H. Liu, Y. Zhang, Y. Ma and F. Mauri, Nature **532** (7597), 81-84 (2016).
14. A. Bianconi and T. Jarlborg, in *Novel Superconducting Materials* (2015), Vol. 1, pp. 37–49.
15. J. E. Hirsch and F. Marsiglio, Physica C: Superconductivity and its Applications **511**, 45-49 (2015).
16. J. E. Hirsch and F. Marsiglio, arXiv:2010.10307v1 [cond-mat.supr-con (2020).
17. A. F. Goncharov, S. S. Lobanov, I. Kruglov, X.-M. Zhao, X.-J. Chen, A. R. Oganov, Z. Konôpková and V. B. Prakapenka, Physical Review B **93** (17), 174105 (2016).
18. T. A. Strobel, P. Ganesh, M. Somayazulu, P. R. C. Kent and R. J. Hemley, Physical Review Letters **107** (25), 255503 (2011).
19. A. F. Goncharov, S. S. Lobanov, V. B. Prakapenka and E. Greenberg, Physical Review B **95** (14), 140101 (2017).
20. B. Guigue, A. Marizy and P. Loubeyre, Physical Review B **95** (2), 020104 (2017).
21. E. J. Pace, X.-D. Liu, P. Dalladay-Simpson, J. Binns, M. Peña-Alvarez, J. P. Attfield, R. T. Howie and E. Gregoryanz, Physical Review B **101** (17), 174511 (2020).
22. M. Einaga, M. Sakata, T. Ishikawa, K. Shimizu, M. I. Eremets, A. P. Drozdov, I. A. Troyan, N. Hirao and Y. Ohishi, Nat Phys **12** (9), 835-838 (2016).
23. H. Nakao, M. Einaga, M. Sakata, M. Kitagaki, K. Shimizu, S. Kawaguchi, N. Hirao and Y. Ohishi, Journal of the Physical Society of Japan **88** (12), 123701 (2019).
24. W. Cui, T. Bi, J. Shi, Y. Li, H. Liu, E. Zurek and R. J. Hemley, Physical Review B **101** (13), 134504 (2020).
25. Y. Sun, Y. Tian, B. Jiang, X. Li, H. Li, T. Iitaka, X. Zhong and Y. Xie, Physical Review B **101** (17), 174102 (2020).
26. See Supplemental Material at http://link.aps.org/supplemental/10.1103/PhysRevLett.xxx for details on Materials and Methods, Tables S1-S6, Figs. S1-S4, and bibliography, which includes Refs. [32-39].
27. D. Duan, Y. Liu, F. Tian, D. Li, X. Huang, Z. Zhao, H. Yu, B. Liu, W. Tian and T. Cui, Sci. Rep. **4**, 6968 (2014).
28. D. Laniel, B. Winkler, E. Bykova, T. Fedotenko, S. Chariton, V. Milman, M. Bykov, V. Prakapenka, L. Dubrovinsky and N. Dubrovinskaia, Physical Review B **102** (13), 134109 (2020).
29. N. P. Salke, M. M. Davari Esfahani, Y. Zhang, I. A. Kruglov, J. Zhou, Y. Wang, E. Greenberg, V. B. Prakapenka, J. Liu, A. R. Oganov and J.-F. Lin, Nature Communications **10** (1), 4453 (2019).
30. A. F. Goncharov, J. H. Eggert, I. I. Mazin, R. J. Hemley and H.-k. Mao, Physical Review B **54** (22), R15590-R15593 (1996).
31. M. S. Somayazulu, L. W. Finger, R. J. Hemley and H. K. Mao, Science **271** (5254), 1400-1402 (1996).

32. V. B. Prakapenka, A. Kubo, A. Kuznetsov, A. Laskin, O. Shkurikhin, P. Dera, M. L. Rivers and S. R. Sutton, High Pressure Research **28** (3), 225-235 (2008).
33. C. Prescher and V. B. Prakapenka, High Pressure Research **35** (3), 223-230 (2015).
34. CrysAlisPro Software System, Rigaku Oxford Diffraction, Oxford, UK. Oxford, UK 2014.
35. G. Sheldrick, Acta Crystallographica Section A **71** (1), 3-8 (2015).
36. G. Sheldrick, Acta Crystallographica Section C **71** (1), 3-8 (2015).
37. O. V. Dolomanov, L. J. Bourhis, R. J. Gildea, J. A. K. Howard and H. Puschmann, Journal of Applied Crystallography **42** (2), 339-341 (2009).
38. N. Holtgrewe, E. Greenberg, C. Prescher, V. B. Prakapenka and A. F. Goncharov, High Pressure Research **39** (3), 457-470 (2019).
39. A. F. Goncharov, P. Beck, V. V. Struzhkin, R. J. Hemley and J. C. Crowhurst, Journal of Physics and Chemistry of Solids **69** (9), 2217-2222 (2008).

Supplemental Material

# Structure and composition of C-S-H compounds up to 143 GPa

Elena Bykova[1], Maxim Bykov[1,2], Stella Chariton[3], Vitali B. Prakapenka[2], Konstantin Glazyrin[4], Andrey Aslandukov[5], Alena Aslandukova[6], Giacomo Criniti[6], Alexander Kurnosov[6], Alexander F. Goncharov[1]

[1] Earth and Planets Laboratory, Carnegie Institution of Washington, 5251 Broad Branch Road NW, Washington, DC 20015, USA

[2] Department of Mathematics, Howard University, Washington, DC 20059, USA

[3] Center for Advanced Radiations Sources, University of Chicago, Chicago, Illinois 60637, USA

[4] Photon Science, Deutsches Elektronen-Synchrotron, Notkestrasse 85, 22607 Hamburg, Germany

[5] Material Physics and Technology at Extreme Conditions, Laboratory of Crystallography, University of Bayreuth, D-95440 Bayreuth, Germany

[6] Bavarian Research Institute of Experimental Geochemistry and Geophysics (BGI), University of Bayreuth, Univesitaetsstrasse 30, 95440 Bayreuth, Germany

**Materials**

The sample synthesis procedure of this work was inspired by that of Ref. [7] aiming to investigate the materials that are similar to those, which show superconductivity at high pressures. In the first batch (B1) of the samples (Table S1), we used a mechanically grinded 1:1 molar mixture of crystalline S (Alfa Aesar, Puratronic®, 99.9995% (metals basis), CAS 7704-34-9) and nano-carbon (Sigma-Aldrich®, >99% (trace metals basis)), CAS 7440-44-0), which was placed in the diamond anvil cell (DAC) cavity and gas loaded by $H_2$ at room temperature to approximately 0.14 GPa. Unlike reported in Ref. [7], the Raman spectra of the synthesized here material did not show any sign of the C-H stretch modes assigned to methane molecules, which were postulated to be the building blocks of the superconducting compound. This inspired us to change the source of carbon and for the sample of the second batch B2 (Table S1), we reacted sulfur with a mixture of $H_2$ and $CH_4$ gases. These were loaded sequentially in the DAC and were allowed to mix in the high-pressure vessel at 0.14 GPa for 40 minutes. The loaded gas mixture was of (1-2.5):1 composition based on the ratio of the intensities of the H-H and C-H stretching modes [31]. For the samples from both batches, we first attempted to employ photochemical synthesis [7] and exposed the C:S mixture to a 488 and 532 nm laser of 10-30 mW power and recorded no change in appearance and Raman spectra at 4 GPa. Instead, we applied up to 180 mW of a focused 458 nm laser, which heated the sample (occasionally we saw melting of S) reaching the synthesis conditions by promoting the chemical reaction with fluid $H_2$ or $CH_4/H_2$ mixture at 4 GPa. Visual observations show that the heated S/C mixtures or S react, and the crystals form slightly away from the laser heating zone, at which moment the laser power was slowly reduced leaving single crystals in the chamber (Fig. S1). In the larger pressure chamber formed by culets with >500 μm diameter, we were not able to grow the crystals by laser heating and instead heated the whole DAC on the hot plate at 200° C for about 2 hours, which resulted in the synthesis of elongated crystals (Fig. S1). We believe that the synthesis of C-S-H crystals at 4 GPa is similar to that of Pace *et al.* [21], who reported the formation of $(H_2S)_2H_2$ crystals from $H_2$ and S in two-step procedure via heat assisted synthesis of $H_2S$.

**Table S1**. Diamond anvil cell experiments.

|  | Reagents | Pressure range (GPa) | Techniques |
|---|---|---|---|
| Batch 1 |  |  |  |
| AG001 | mixed C & S gas loaded with $H_2$ | 4.0 - 5.3 | XRD/Raman |
| AG002 | mixed C & S gas loaded with $H_2$ | 4 - 45 | Raman/abs. |
| AG003 | mixed C & S gas loaded with $H_2$ | 4 - 58 | XRD/Raman |
| Batch 2 |  |  |  |
| AG004 | S gas loaded with mixed $H_2$ & $CH_4$ | 4 - 65 | XRD/Raman/abs. |
| AG005 | S gas loaded with mixed $H_2$ & $CH_4$ | 4 - 143 | XRD/Raman/abs. |

abs. – absorption; XRD - X-ray diffraction

**Experimental methods**

**1. X-ray diffraction (XRD) data collection**

The XRD measurements were conducted at the 13-IDD beamline at the Advanced Photon Source (APS), Chicago, USA (Pilatus CdTe 1M detector, λ = 0.29520 Å, KB-mirror focusing) and at the Extreme Conditions Beamline P02.2, DESY, Hamburg, Germany (Perkin Elmer XRD1621 flat panel detector, λ = 0.2908 Å, KB-mirror focusing). Sample-to-detector distance, coordinates of the beam center, tilt angle and tilt plane rotation angle of the detector images were calibrated using $LaB_6$ (13-IDD experiments) or $CeO_2$ (P02.2 experiments) powders. The single-crystal XRD images were recorded while rotating the sample about a single ω-axis from -30 to +30° in small steps of 0.5°.

Laser-heating experiments were carried out on a state-of-the art stationary double-side laser-heating set-up installed at IDD-13 [32] and P02.2 beamlines. The temperature was measured by the standard spectroradiometry method. Here, we report the room temperature XRD data.

**2. XRD data processing**

DIOPTAS software[33] was used for preliminary analysis and for integration of the 2-dimentional images to 1-dimentional diffraction patterns.

Processing of single-crystal XRD data (the unit cell determination and integration of the reflection intensities) was performed using CrysAlisPro software [34]. Empirical absorption correction was applied using spherical harmonics, implemented in the SCALE3 ABSPACK scaling algorithm, which is included in the CrysAlisPro software. A single crystal of an orthoenstatite (($Mg_{1.93}$,$Fe_{0.06}$)($Si_{1.93}$,$Al_{0.06}$)$O_6$, *Pbca*, *a* = 18.2391(3), *b* = 8.8117(2), *c* = 5.18320(10) Å), was used to calibrate instrument model of CrysAlisPro software (sample-to-detector distance, the detector's origin, offsets of the goniometer angles and rotation of the X-ray beam and the detector around the instrument axis). The calibration crystal was measured in a diamond anvil cell without pressure medium in similar experimental conditions as the studied samples (narrow slicing mode, rotation about ω-axis from -30 to +30° with 0.5°step size).

**3. Structure solution and refinement.**

The structure solution was performed by method of intrinsic phasing implemented in SHELXT [35]. The crystal structures were refined against $F^2$ on all data by full-matrix least squares with the SHELXL [36] software. SHELXT and SHELXL programs are implemented in Olex2 software package [37].

*I4/mcm structures.* Structure solution yields only positions of sulfur atoms. Hydrogen molecules were found after analysis of the residual density maps: they describe the highest residual density peaks. Additionally we used constraints on the H-H distances (0.75±0.02 Å according to Duan *et al.* [27]). We could not succeed in finding positions of hydrogen atoms attached to sulfur, probably due to their strong disorder and due to the incompleteness of the datasets. For the B1 and B2 samples, no residual density peaks were found, which could correspond to a possible isolated $CH_4$ molecule. The hypothesis of a random substitution of sulfur by carbon does not improve the refinement, since the structure has only one position of a heavy atom and its refined occupancy is directly correlated with a scale factor. Therefore, carbon was not introduced into the refinement. Thermal parameters of sulfur were refined in anisotropic approximation, hydrogen thermal parameters were refined only in case of sufficient data/parameter ratio and were fixed to 0.05 Å$^2$ otherwise.

Unit cell parameters, structural data and details of the structure refinements of B1 and B2 samples are given in Supplementary Tables S2-S4.

*C/2c structure at 27 GPa (AG003).*

At 27 GPa additional reflections were observed nearby the reflections belonging to the *I4/mcm* phase. Further examination of the reciprocal space has shown that the peaks belong to two twin domains that partly overlap each other and with the *I4/mcm* phase as well. The new phase was indexed in monoclinic *C*-centered unit cell with parameters $a$ = 8.255(16), $b$ = 6.3944(18), $c$ = 12.14(3) Å, β = 99.3(3)°. The twin law is (0.5826 -0.7555 -0.3315  0.0744 0.7445 -0.3344 1.0895 0.7712 0.6691) which corresponds to a rotation by about 60 degrees around [0.58, -0.58, 0.57] direction in the reciprocal space. Due to multiple overlaps between reflections, only sulfur positions could be determined reliably. The final $R_1$ remains relatively high (~13%), therefore no definite conclusions could be made about positions of hydrogens. Unit cell parameters, structural data and details of the structure refinements are given in Supplementary Table S5.

*Pnma structure at 143 GPa (AG005).* Similarly to the *I*4/*mcm* structures only sulfur positons could be determined after the structure solution. Thermal parameters of sulfur atoms were refined in anisotropic approximation. On the residual density maps, there are several residual peaks of 1-0.5 e/A$^3$, but among them no appropriate ones were found to reasonably describe an isolated $CH_4$ molecule, therefore carbon likely remains on the sulfur positions. Similarly to the *I*4/*mcm* structures, occupancy of C would be coupled with the scale factor, therefore carbon was not introduced into the refinement. No definite conclusions could be made about positions of hydrogens.

Unit cell parameters, structural data and details of the structure refinements are given in Supplementary Table S6 and Fig. S2.

## 4. Raman spectroscopy

Our custom made DAC optimized confocal Raman systems at the EPL and GSECARS feature narrow line excitation wavelength 473, 488, 532, and 660 nm, ultralow frequency (<10 cm$^{-1}$) Raman filters, and single grating spectrographs with deep-depleted back illuminated CCD detectors[38].

## 5. Optical spectroscopy (absorption measurements)

Our custom made all mirror system for optical spectroscopy system [39] has been adapted to incorporate a supercontinuum (white laser) light source, which was delivered via and optical fiber. The light spot diameter at the sample position was about 5 µm in diameter permitting to perform spatially resolved measurements of our DAC samples (Fig. S1).

| Table S2. Details of crystal structure refinements for B1 samples at high pressures* ||||||||
|---|---|---|---|---|---|---|
| Sample name | AG001_p01_s1 | AG003_p01_s2 | AG003_p02_s2** | AG003_p03_s1 | AG003_p04_s2 | AG003_p05_s1** |
| Pressure, GPa | **5.3(4)** | **16.9(5)** | **27.8(7)** | **34.8(8)** | **44.6(8)** | **57.9(9)** |
| $a$ (Å) | 7.2841(13) | 6.638(3) | 6.328(3) | 6.1498(12) | 6.023(5) | 5.870(3) |
| $c$ (Å) | 6.0053(15) | 5.487(4) | 5.254(2) | 5.1648(10) | 5.017(4) | 4.887(2) |
| $V$ (Å$^3$) | 318.63(14) | 241.7(3) | 210.4(2) | 195.33(9) | 182.0(3) | 168.41(17) |
| Reflections collected | 250 | 254 | | 75 | 104 | |
| Independent reflections | 132 | 99 | | 34 | 37 | |
| Independent reflections [$I > 2\sigma(I)$] | 120 | 72 | | 31 | 36 | |
| Refined parameters | 7 | 6 | | 6 | 6 | |
| $R_{int}(F^2)$ | 0.0163 | 0.0225 | | 0.0273 | 0.1207 | |
| $R(\sigma)$ | 0.0209 | 0.0189 | | 0.0226 | 0.0457 | |
| $R_1$ [$I > 2\sigma(I)$] | 0.0451 | 0.0784 | | 0.0769 | 0.0744 | |
| $wR_2$ [$I > 2\sigma(I)$] | 0.1254 | 0.1953 | | 0.228 | 0.213 | |
| $R_1$ | 0.0481 | 0.0917 | | 0.0782 | 0.0748 | |
| $wR_2$ | 0.1293 | 0.2132 | | 0.2302 | 0.2133 | |
| Goodness of fit on $F^2$ | 1.067 | 1.142 | | 1.396 | 1.195 | |
| $\Delta\rho_{max}$ ($e$ / Å$^3$) | 0.299 | 0.824 | | 0.658 | 0.571 | |
| $\Delta\rho_{min}$ ($e$ / Å$^3$) | -0.614 | -0.742 | | -0.703 | -0.914 | |
| $x$(S1) | 0.16307(8) | 0.1668(3) | | 0.1747(4) | 0.1722(4) | |
| $y$(H2) | 0 | 0.0565(16) | | 0.0610(18) | 0.0622(18) | |
| $z$(H2) | 0.1885(17) | 0.25 | | 0.25 | 0.25 | |
| $U_{eq}$(S1) (Å$^2$)*** | 0.0338(5) | 0.0385(10) | | 0.028(3) | 0.032(3) | |
| $U_{iso}$(H2) (Å$^2$) | 0.050(18) | 0.05 | | 0.05 | 0.05 | |
| $U_{11}$(S1) (Å$^2$) | 0.0346(6) | 0.0437(12) | | 0.031(4) | 0.034(4) | |
| $U_{22}$(S1) (Å$^2$) | 0.0346(6) | 0.0437(12) | | 0.031(4) | 0.034(4) | |
| $U_{33}$(S1) (Å$^2$) | 0.0322(6) | 0.0281(15) | | 0.023(4) | 0.026(3) | |
| $U_{12}$(S1) (Å$^2$) | 0.0002(4) | -0.0008(8) | | 0.0018(16) | 0.0005(14) | |
| $d1$(S1…S1) (Å) | 3.3597(18) | 3.132(6) | | 3.039(8) | 2.934(7) | |
| $d2$(S1…S1) (Å) | 3.4962(11) | 3.157(3) | | 2.895(3) | 2.837(4) | |
| $d3$(S1…S1) (Å) | 3.8288(10) | 3.526(3) | | 3.360(3) | 3.255(4) | |
| $d4$(S1…S1) (Å) | 3.8559(8) | 3.498(2) | | 3.2112(16) | 3.154(3) | |
| Data collection | APS, 13-IDD beamline, Pilatus CdTe 1M detector, $\lambda$ = 0.29520 Å ||||||

*(H$_2$(S,C))$_2$H$_2$ adopts Al$_2$Cu crystal structure, space group $I4/mcm$, $Z = 4$,

atoms' Wyckoff positions (below 10 GPa): S1 8$h$ ($x$, $x$ + 0.5, 0)
                                     H2 8$f$ (0, 0, $z$)

atoms' Wyckoff positions (above 10 GPa): S1 8$h$ ($x$, $x$ + 0.5, 0)
                                     H2 16$j$ (0, $y$, 0.25)

** Structure refinement was not possible
***$U_{eq}$ is defined as one third of the trace of the orthogonalized $U^{ij}$ tensor.

| Table S3. Details of crystal structure refinements for B2-AG004 samples at high pressures* | | | | | | | |
|---|---|---|---|---|---|---|---|
| Sample name | AG004_p01_s1 | AG004_p02_s1 | AG004_p03_s1 | AG004_p04_s1** | AG004_p05_s1 | AG004_p06_s1 | AG004_p07_s1 |
| Pressure, GPa | 8.6(2) | 25.5(2) | 35.4(2) | 42.3(3) | 49.2(3) | 54(1) | 64(3) |
| $a$ (Å) | 6.9803(7) | 6.435(2) | 6.234(3) | 6.1487(12) | 6.0439(16) | 5.9864(9) | 5.857(2) |
| $c$ (Å) | 5.7811(9) | 5.338(4) | 5.168(4) | 5.0852(17) | 4.9674(15) | 4.9133(11) | 4.836(4) |
| $V$ (Å$^3$) | 281.68(7) | 221.1(2) | 200.8(2) | 192.25(8) | 181.45(11) | 176.08(7) | 165.92(18) |
| Reflections collected | 301 | 196 | 188 |  | 172 | 370 | 343 |
| Independent reflections | 129 | 95 | 91 |  | 84 | 153 | 140 |
| Independent reflections [$I > 2\sigma(I)$] | 122 | 78 | 79 |  | 78 | 136 | 113 |
| Refined parameters | 7 | 6 | 6 |  | 6 | 6 | 6 |
| $R_{int}(F^2)$ | 0.0045 | 0.0194 | 0.0155 |  | 0.0154 | 0.0285 | 0.0407 |
| $R(\sigma)$ | 0.0049 | 0.0155 | 0.0142 |  | 0.0122 | 0.0304 | 0.0529 |
| $R_1$ [$I > 2\sigma(I)$] | 0.0279 | 0.0776 | 0.0737 |  | 0.0574 | 0.0632 | 0.0825 |
| $wR_2$ [$I > 2\sigma(I)$] | 0.0828 | 0.185 | 0.1839 |  | 0.1624 | 0.1612 | 0.2082 |
| $R_1$ | 0.0287 | 0.0831 | 0.0783 |  | 0.0581 | 0.0656 | 0.0978 |
| $wR_2$ | 0.0833 | 0.1953 | 0.1948 |  | 0.1634 | 0.1641 | 0.2404 |
| Goodness of fit on $F^2$ | 1.241 | 1.098 | 1.14 |  | 1.201 | 1.146 | 1.108 |
| $\Delta\rho_{max}(e / Å^3)$ | 0.378 | 1.114 | 1.003 |  | 1.034 | 1.446 | 1.375 |
| $\Delta\rho_{min}(e / Å^3)$ | -0.316 | -0.891 | -0.952 |  | -0.757 | -0.901 | -1.353 |
| $x$(S1) | 0.16390(5) | 0.1664(2) | 0.16767(17) |  | 0.16834(11) | 0.16871(8) | 0.16986(14) |
| $y$(H2) | 0 | 0.0581(16) | 0.0600(17) |  | 0.0619(17) | 0.0626(17) | 0.0639(17) |
| $z$(H2) | 0.1866(17) | 0.25 | 0.25 |  | 0.25 | 0.25 | 0.25 |
| $U_{eq}$(S1) (Å$^2$)*** | 0.0270(3) | 0.0255(9) | 0.0220(8) |  | 0.0148(7) | 0.0126(4) | 0.0138(6) |
| $U_{iso}$(H2) (Å$^2$) | 0.047(13) | 0.05 | 0.05 |  | 0.05 | 0.05 | 0.05 |
| $U_{11}$(S1) (Å$^2$) | 0.0271(4) | 0.0280(10) | 0.0246(10) |  | 0.0165(8) | 0.0142(5) | 0.0158(8) |
| $U_{22}$(S1) (Å$^2$) | 0.0271(4) | 0.0280(10) | 0.0246(10) |  | 0.0165(8) | 0.0142(5) | 0.0158(8) |
| $U_{33}$(S1) (Å$^2$) | 0.0269(4) | 0.0207(11) | 0.0166(11) |  | 0.0115(9) | 0.0092(5) | 0.0096(8) |
| $U_{12}$(S1) (Å$^2$) | 0.0000(2) | 0.0001(6) | 0.0003(6) |  | -0.0007(4) | 0.0002(2) | 0.0006(4) |
| $d1$(S1…S1) (Å) | 3.2359(11) | 3.029(4) | 2.956(3) |  | 2.878(2) | 2.8565(13) | 2.814(3) |
| $d2$(S1…S1) (Å) | 3.3534(7) | 3.072(3) | 2.964(2) |  | 2.8491(12) | 2.8160(8) | 2.758(2) |
| $d3$(S1…S1) (Å) | 3.6866(6) | 3.422(2) | 3.324(2) |  | 3.2108(11) | 3.1804(7) | 3.1315(19) |
| $d4$(S1…S1) (Å) | 3.6913(4) | 3.3928(15) | 3.2814(16) |  | 3.1791(9) | 3.1475(6) | 3.0755(13) |
| Data collection | (1) | (1) | (1) | (1) | (1) | (2) | (2) |

*(H$_2$(S,C))$_2$H$_2$ adopts Al$_2$Cu crystal structure, space group $I4/mcm$, $Z = 4$,
atoms' Wyckoff positions (below 10 GPa):   S1 8$h$ ($x$, $x$ + 0.5, 0)
    H2 8$f$ (0, 0, $z$)

atoms' Wyckoff positions (above 10 GPa):   S1 8$h$ ($x$, $x$ + 0.5, 0)
    H2 16$j$ (0, $y$, 0.25)

(1) APS, 13-IDD beamline, Pilatus CdTe 1M detector, $\lambda$ = 0.29520 Å
(2) DESY, P02.2. ECB beamline, Perkin Elmer detector, $\lambda$ = 0.3344 Å

** Structure refinement was not possible
***$U_{eq}$ is defined as one third of the trace of the orthogonalized $U^{ij}$ tensor.

| Table S4. Details of crystal structure refinements for B2-AG005 samples at high pressures* | | | | | | | |
|---|---|---|---|---|---|---|---|
| Sample name | AG005_p01_s1 | AG005_p02_s1 | AG005_p03_s1 | AG005_p04_s1 | AG005_p05_s1 | AG005_p06_s1 | AG005_p07_s1** |
| Pressure, GPa | 14.3(4) | 71(1) | 81(1) | 98(2) | 118.5(3) | 137(3) | 143(3) |
| $a$ (Å) | 6.732(2) | 5.812(3) | 5.714(5) | 5.624(3) | 5.512(4) | 5.411(4) | 5.391(6) |
| $c$ (Å) | 5.6015(17) | 4.7794(16) | 4.717(3) | 4.631(3) | 4.497(3) | 4.383(3) | 4.374(4) |
| $V$ (Å$^3$) | 253.84(18) | 161.46(18) | 154.0(3) | 146.48(19) | 136.6(2) | 128.31(19) | 127.1(3) |
| Reflections collected | 254 | 136 | 137 | 108 | 124 | 122 | |
| Independent reflections | 81 | 52 | 49 | 41 | 37 | 41 | |
| Independent reflections [$I > 2\sigma(I)$] | 72 | 48 | 46 | 39 | 34 | 37 | |
| Refined parameters | 7 | 6 | 6 | 6 | 6 | 6 | |
| $R_{int}(F^2)$ | 0.018 | 0.0179 | 0.0075 | 0.0206 | 0.0297 | 0.0098 | |
| $R(\sigma)$ | 0.0118 | 0.0107 | 0.0049 | 0.0103 | 0.019 | 0.0052 | |
| $R_1$ [$I > 2\sigma(I)$] | 0.0281 | 0.0587 | 0.0492 | 0.0646 | 0.0611 | 0.0722 | |
| $wR_2$ [$I > 2\sigma(I)$] | 0.0778 | 0.1508 | 0.1155 | 0.1621 | 0.1763 | 0.1632 | |
| $R_1$ | 0.0299 | 0.0616 | 0.0519 | 0.0653 | 0.0626 | 0.0742 | |
| $wR_2$ | 0.0797 | 0.1522 | 0.1172 | 0.1683 | 0.178 | 0.1645 | |
| Goodness of fit on $F^2$ | 1.203 | 1.28 | 1.308 | 1.385 | 1.348 | 1.238 | |
| $\Delta\rho_{max}$ ($e$ / Å$^3$) | 0.254 | 0.778 | 0.553 | 0.795 | 0.615 | 1.127 | |
| $\Delta\rho_{min}$ ($e$ / Å$^3$) | -0.256 | -0.407 | -0.512 | -0.546 | -0.519 | -0.822 | |
| $x$(S1) | 0.16517(7) | 0.1718(2) | 0.17242(16) | 0.1739(2) | 0.1761(4) | 0.1767(3) | |
| $y$(H2) | 0.0556(15) | 0.0646(18) | 0.0656(19) | 0.0665(19) | 0.068(2) | 0.069(2) | |
| $z$(H2) | 0.25 | 0.25 | 0.25 | 0.25 | 0.25 | 0.25 | |
| $U_{eq}$(S1) (Å$^2$)*** | 0.0227(4) | 0.0199(9) | 0.0180(8) | 0.0281(14) | 0.0296(15) | 0.0214(11) | |
| $U_{iso}$(H2) (Å$^2$) | 0.040(18) | 0.05 | 0.05 | 0.05 | 0.05 | 0.05 | |
| $U_{11}$(S1) (Å$^2$) | 0.0226(5) | 0.0234(12) | 0.0209(10) | 0.0298(16) | 0.0372(18) | 0.0259(15) | |
| $U_{22}$(S1) (Å$^2$) | 0.0226(5) | 0.0234(12) | 0.0209(10) | 0.0298(16) | 0.0372(18) | 0.0259(15) | |
| $U_{33}$(S1) (Å$^2$) | 0.0230(5) | 0.0131(12) | 0.0121(9) | 0.0247(19) | 0.015(2) | 0.0124(17) | |
| $U_{12}$(S1) (Å$^2$) | -0.0004(4) | 0.0001(10) | 0.0014(7) | -0.0009(8) | 0.0047(19) | 0.0021(10) | |
| $d1$(S1…S1) (Å) | 3.1449(17) | 2.824(4) | 2.786(3) | 2.767(4) | 2.745(7) | 2.704(5) | |
| $d2$(S1…S1) (Å) | 3.2331(11) | 2.7138(19) | 2.6710(18) | 2.613(2) | 2.527(3) | 2.462(3) | |
| $d3$(S1…S1) (Å) | 3.5762(10) | 3.1141(19) | 3.0731(19) | 3.031(2) | 2.970(3) | 2.908(3) | |
| $d4$(S1…S1) (Å) | 3.5543(12) | 3.0450(18) | 2.991(3) | 2.9395(18) | 2.874(3) | 2.819(2) | |
| Data collection | APS, 13-IDD beamline, Pilatus CdTe 1M detector, $\lambda$ = 0.29520 Å | | | | | | |

*(H$_2$(S,C))$_2$H$_2$ adopts Al$_2$Cu crystal structure, space group $I4/mcm$, $Z = 4$,
atoms' Wyckoff positions (below 10 GPa): S1 8$h$ ($x$, $x$ + 0.5, 0)
H2 8$f$ (0, 0, $z$)

atoms' Wyckoff positions (above 10 GPa): S1 8$h$ ($x$, $x$ + 0.5, 0)
H2 16$j$ (0, $y$, 0.25)

** Structure refinement was not possible
***$U_{eq}$ is defined as one third of the trace of the orthogonalized $U^{ij}$ tensor.

**Table S5. Details of crystal structure refinement for C2/c phase at 27 GPa**

| Sample name | AG003_p02_s2 |
|---|---|
| Pressure, GPa | 24.6(3) |
| $a$ (Å) | 8.255(16) |
| $b$ (Å) | 6.3944(18) |
| $c$ (Å) | 12.14(3) |
| $\beta$ (°) | 99.3(3) |
| $V$ (Å$^3$) | 632(2) |
| $Z$ | 4 |
| Reflections collected | 696 |
| Independent reflections | 358 |
| Independent reflections [$I > 2\sigma(I)$] | 266 |
| Refined parameters | 28 |
| $R_{int}(F^2)$ | 0.0786 |
| $R(\sigma)$ | 0.0751 |
| $R_1$ [$I > 2\sigma(I)$] | 0.1279 |
| $wR_2$ [$I > 2\sigma(I)$] | 0.3195 |
| $R_1$ | 0.1498 |
| $wR_2$ | 0.3392 |
| Goodness of fit on $F^2$ | 1.329 |
| $\Delta\rho_{max}$ ($e$ / Å$^3$) | 1.219 |
| $\Delta\rho_{min}$ ($e$ / Å$^3$) | -1.226 |
| Data collection | APS, 13-IDD beamline, Pilatus CdTe 1M detector, $\lambda$ = 0.29520 Å |

**Atomic coordinates and equivalent displacement parameters $U_{eq}$***

| Atom | Wyckoff position | $x$ | $y$ | $z$ | $U_{eq}$ (Å$^2$) |
|---|---|---|---|---|---|
| S1 | 8$f$ | 0.1540(7) | 0.0826(4) | 0.4448(5) | 0.020(4) |
| S2 | 8$f$ | 0.1957(7) | 0.0608(4) | 0.7222(5) | 0.024(4) |
| S3 | 8$f$ | 0.0309(6) | 0.3899(4) | 0.8883(5) | 0.017(4) |

**Anisotropic displacement parameters, in Å$^2$**

| Atom | $U_{11}$ | $U_{22}$ | $U_{33}$ | $U_{12}$ | $U_{13}$ | $U_{23}$ |
|---|---|---|---|---|---|---|
| S1 | 0.016(7) | 0.0094(11) | 0.041(8) | -0.0015(15) | 0.021(7) | -0.0002(16) |
| S2 | 0.020(7) | 0.0082(12) | 0.051(9) | -0.0003(14) | 0.029(8) | 0.0004(15) |
| S3 | 0.010(7) | 0.0092(13) | 0.037(9) | 0.0006(12) | 0.018(8) | 0.0007(14) |

*$U_{eq}$ is defined as one third of the trace of the orthogonalized $U^{ij}$ tensor.

**Table S6. Details of crystal structure refinement for *Pnma* phase at 143 GPa**

| Sample name | AG005_p07_q1_s1 |
|---|---|
| Pressure, GPa | 143(2) |
| *a* (Å) | 7.591(5) |
| *b* (Å) | 4.439(13) |
| *c* (Å) | 7.641(6) |
| *V* (Å$^3$) | 257.5(8) |
| *Z* | 4 |
| Reflections collected | 363 |
| Independent reflections | 237 |
| Independent reflections [$I > 2\sigma(I)$] | 153 |
| Refined parameters | 25 |
| $R_{int}(F^2)$ | 0.0342 |
| $R(\sigma)$ | 0.0486 |
| $R_1$ [$I > 2\sigma(I)$] | 0.087 |
| $wR_2$ [$I > 2\sigma(I)$] | 0.2349 |
| $R_1$ | 0.1244 |
| $wR_2$ | 0.2686 |
| Goodness of fit on $F^2$ | 1.147 |
| $\Delta\rho_{max}(e / Å^3)$ | 1.065 |
| $\Delta\rho_{min}(e / Å^3)$ | -1.058 |
| Data collection | APS, 13-IDD beamline, Pilatus CdTe 1M detector, λ = 0.29520 Å |

**Atomic coordinates and equivalent displacement parameters $U_{eq}$***

| Atom | Wyckoff position | *x* | *y* | *z* | $U_{eq}$ (Å$^2$) |
|---|---|---|---|---|---|
| S1 | 4*c* | 0.2347(5) | 0.75 | 0.1517(5) | 0.0175(17) |
| S2 | 4*c* | 0.0736(5) | 0.75 | 0.4930(4) | 0.0151(16) |
| S3 | 4*c* | 0.2383(5) | 0.75 | 0.8126(5) | 0.0170(17) |
| S4 | 4*c* | 0.0589(5) | 0.25 | 0.9872(5) | 0.0152(17) |

**Anisotropic displacement parameters, in Å$^2$**

| Atom | $U_{11}$ | $U_{22}$ | $U_{33}$ | $U_{12}$ | $U_{13}$ | $U_{23}$ |
|---|---|---|---|---|---|---|
| S1 | 0.0189(17) | 0.025(6) | 0.0082(17) | 0 | 0.0011(12) | 0 |
| S2 | 0.0164(14) | 0.023(6) | 0.0061(17) | 0 | -0.0012(11) | 0 |
| S3 | 0.0167(15) | 0.029(6) | 0.0055(15) | 0 | 0.0018(12) | 0 |
| S4 | 0.0176(15) | 0.022(6) | 0.0062(16) | 0 | 0.0009(11) | 0 |

*$U_{eq}$ is defined as one third of the trace of the orthogonalized $U^{ij}$ tensor.

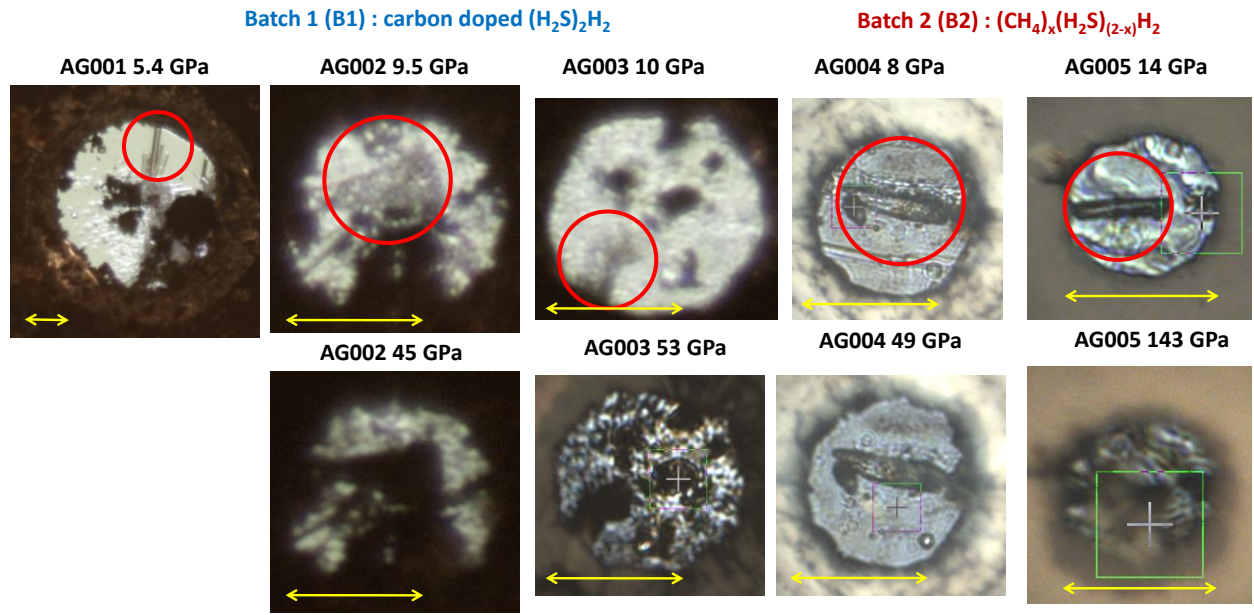

**Figure S1**. Microphotograph of C-S-H samples examined in this work at different pressures. The synthesized crystals are marked by red circles. Other pieces in the chamber for the B1 samples are unreacted mixtures of carbon and sulfur and ruby balls to measure pressure. The lengths of yellow bidirectional arrows correspond to 50 μm.

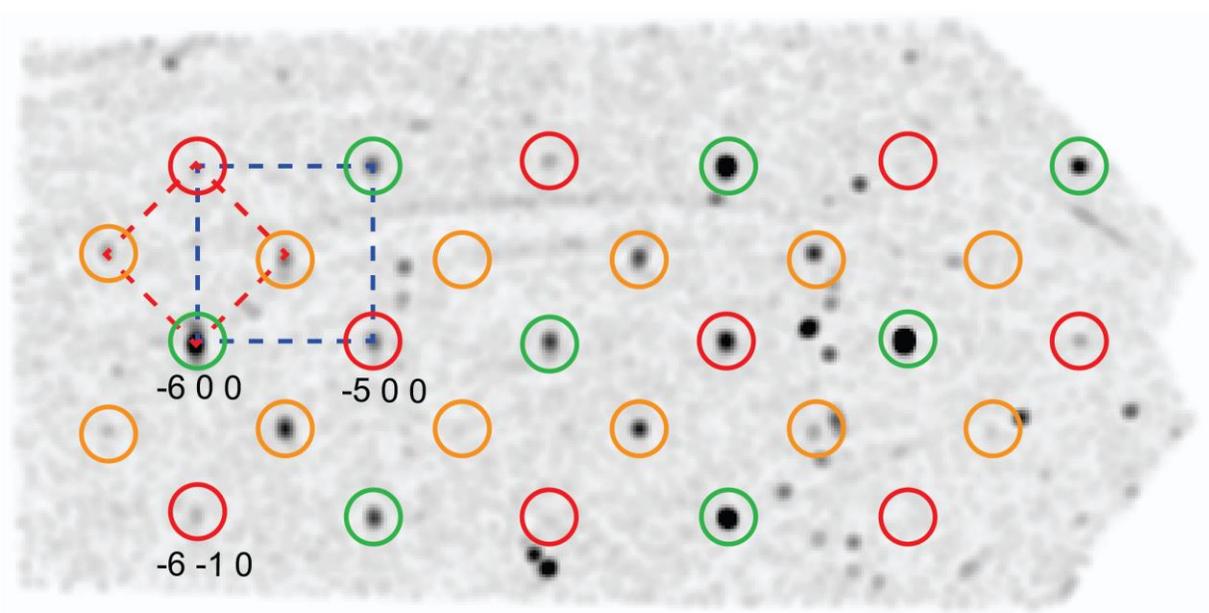

**Figure S2.** Reciprocal space reconstruction of (*hk0*) layer for the sample AG005 after the laser heating at 138.3 GPa: green circles – reflections that follow *I*-centering in the parent *I*4/*mcm* structure (blue dash line); red circles – reflections violating *I*-centering; yellow circles – satellite reflections at $0.5(a^*+ b^*)$. All reflections can be indexed in an orthorhombic *P*-centered unit cell with $a = 4.438945$, $b = 7.660409$, $c = 7.559471$ Å (red dash line).

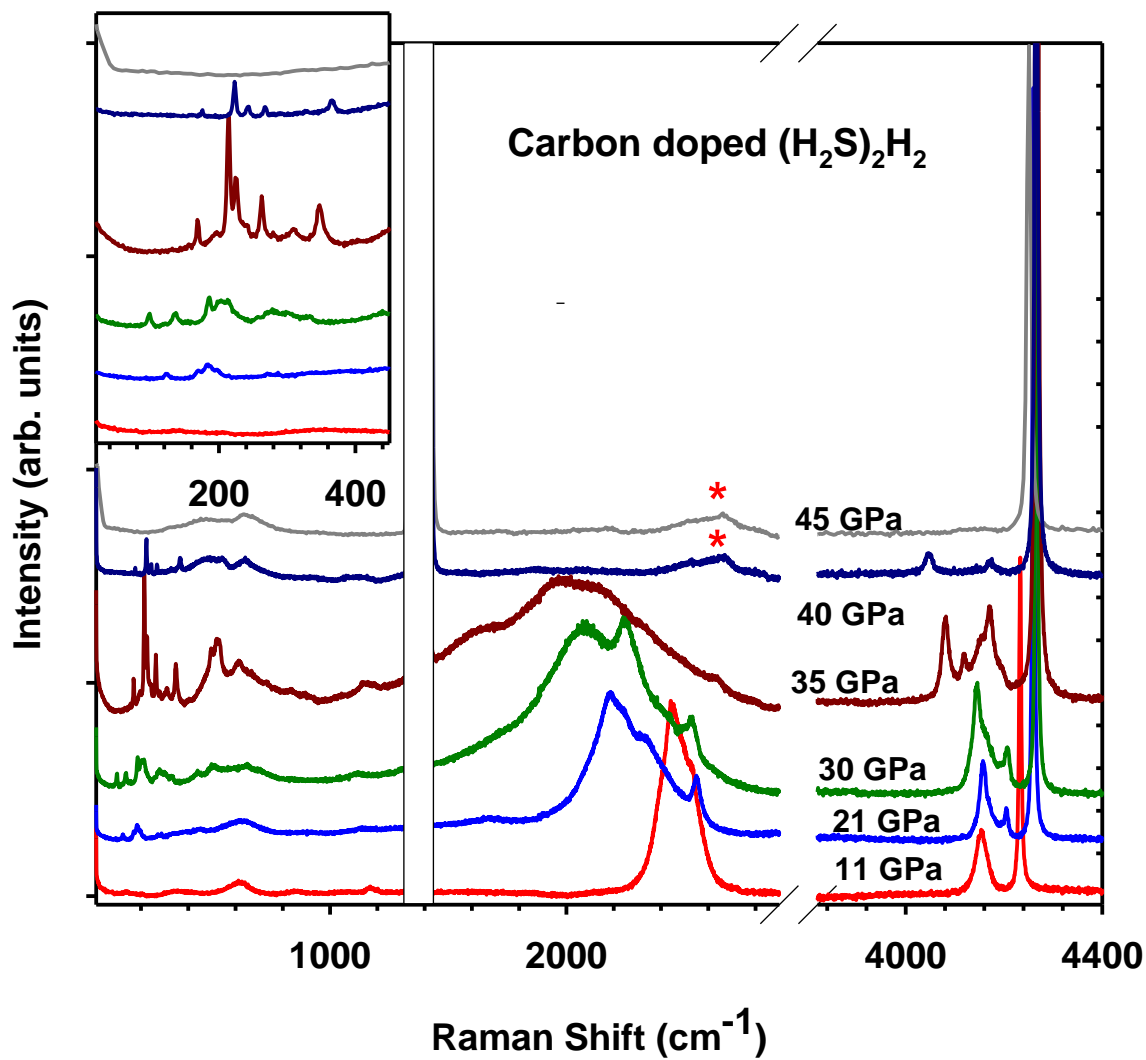

**Figure S3**. Raman spectra of the B1 batch AG002 sample as a function of pressure. The inset zoom in on a low-frequency range of the lattice modes. The first-order diamond signal of diamond anvils is masked by a rectangle, while the second order diamond anvil signal at 40 and 45 GPa is marked by an asterisk.

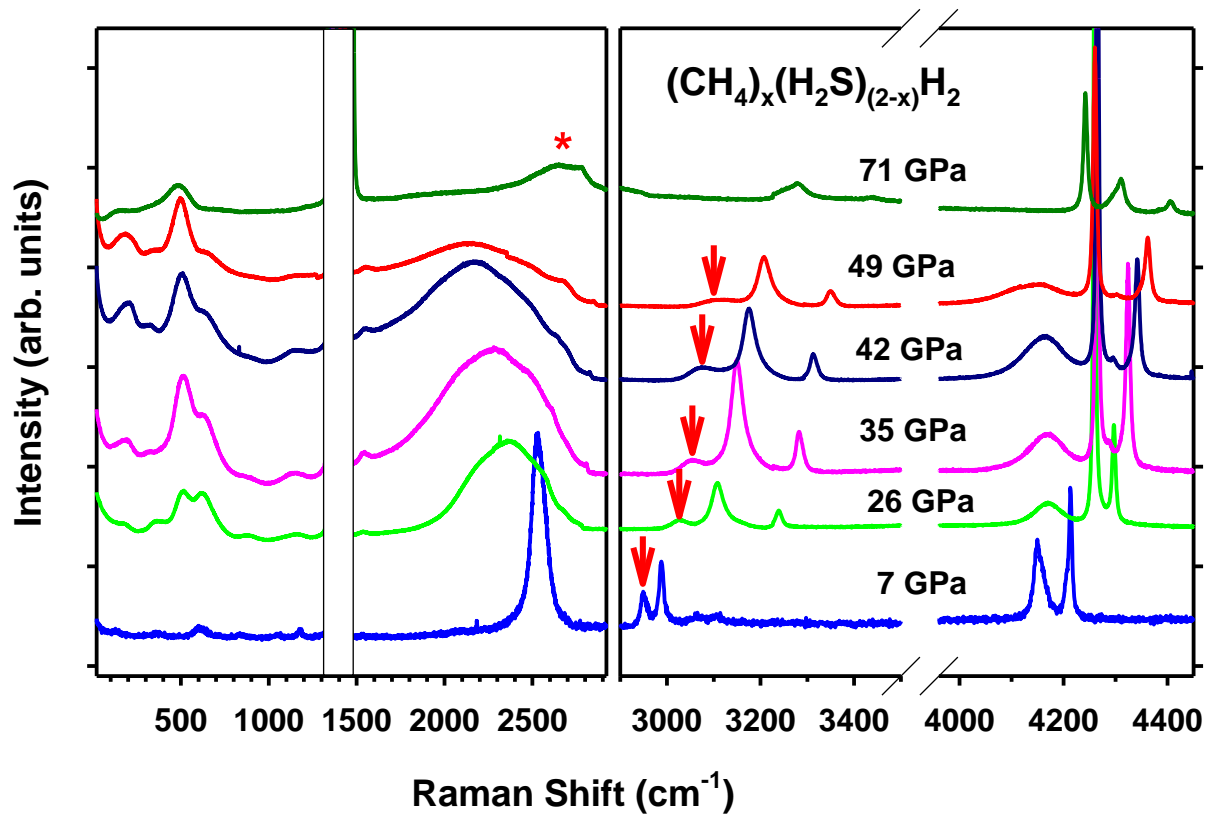

**Figure S4**. Raman spectra of the B2 batch AG004(5) samples as a function of pressure. Red arrows mark the C-H stretching band peaks of methane molecules, which are red shifted with respect to those of a $CH_4$-$H_2$ compound surrounding them. The first-order diamond signal of diamond anvils is masked by a rectangle, while the second order diamond anvil signal at 71 GPa is marked by an asterisk.